\documentclass[aps,pre,twocolumn,groupedaddress,showpacs,floatfix,amsmath,nofootinbib]{revtex4}
\usepackage{graphicx,amsmath}
\begin{document}
\preprint{}
\title{Growing length scale accompanying the vitrification: \\ 
A perspective based on non-singular density fluctuations}  
\author{Akira Furukawa}
\email{furu@iis.u-tokyo.ac.jp}
\affiliation{Institute of Industrial Science, University of Tokyo, Meguro-ku, Tokyo 153-8505, Japan}
\date{\today}
\begin{abstract}
In glass forming liquids close to the glass transition point, even a very slight increase in the {\it macroscopic} density results in a dramatic slowing down of the {\it macroscopic} relaxation. Concomitantly, the {\it local} density itself fluctuates in space. Therefore, one can imagine that even very small {\it local} density variations control the {\it local} glassy nature. Based on this perspective, a model for describing growing length scale accompanying the vitrification is introduced, in which we assume that in a subsystem whose density is above a certain threshold value, $\rho_{\rm c}$, owing to steric constraints, particle rearrangements are highly suppressed for a sufficiently long time period ($\sim$ structural relaxation time). We regard such a subsystem as a glassy cluster. Then, based on the statistics of the subsystem-density, we predict that with compression (increasing average density $\rho$) at a fixed temperature $T$ in supercooled states, the characteristic length of the clusters, $\xi$, diverges as $\xi\sim(\rho_{\rm c}-\rho)^{-2/d}$, where $d$ is the spatial dimensionality. This $\xi$ measures the average persistence length of the steric constraints in blocking the rearrangement motions and is determined by the subsystem density. Additionally, with decreasing $T$ at a fixed $\rho$, the length scale diverges in the same manner as $\xi\sim(T-T_{\rm c})^{-2/d}$, for which $\rho$ is identical to $\rho_{\rm c}$ at $T=T_{\rm c}$. The exponent describing the diverging length scale is the same as the one predicted by some theoretical models and indeed has been observed in some simulations and experiments. However, the basic mechanism for this divergence is different; that is, we do not invoke thermodynamic anomalies associated with the thermodynamic phase transition as the origin of the growing length scale. We further present arguements for the cooperative properties of the structural relaxation based on the clusters. 
\end{abstract}
\maketitle
\section{Introduction}
As a supercooled liquid approaches the glass transition point, the structural relaxation slows dramatically, and the viscosity increases steeply. The origin of this viscous slowdown remains the central issue in glass physics \cite{Ediger-Angell-Nagel,Sillescu,Ediger,Debenedetti-Stillinger,DyreR,Binder-KobB,Berchier-BiroliR,dynamic_heterogeneityB,Tanaka_review,Sastry_review}. The general dynamic features of the glass transition process appear to be as follows:
 
(1) In the normal liquid regime far above the glass transition point, particle motions and the resultant structural relaxation dynamics are spatially uncorrelated.

(2) By increasing the density $\rho$ at a fixed temperature $T$ (decreasing $T$ at a fixed $\rho$), a crossover from the normal to supercooled state occurs gradually around a certain density $\rho_{\rm 0}$ (temperature $T_{\rm 0}$). For $\rho\gtrsim \rho_{\rm 0}$ ($T\lesssim T_{\rm 0}$), the motion of a particle is hindered by its neighbors, which is the so-called caging. 
Then, in the supercooled liquid regime, structural rearrangements occur cooperatively; the more glassy the system is, the slower and more cooperative the dynamics, and the larger the cooperative or correlation length \cite{Berchier-BiroliR,dynamic_heterogeneityB,Tanaka_review,Sastry_review}. 

(3) With further compression (cooling), the system is vitrified at the glass transition density $\rho_{\rm g}$ (glass transition temperature $T_{\rm g}$), where the rearrangement motions are almost frozen. 

Following the seminal work of Adam and Gibbs \cite{Adam_Gibbs}, many theoretical and numerical efforts have been attempted to elucidate the underlying mechanism of the correlation or cooperative structural rearrangements, in recent years, particularly from thermodynamic perspectives (see reviews \cite{Berchier-BiroliR,dynamic_heterogeneityB,Tanaka_review,Sastry_review}). Although such approaches are appealing, it is an open question whether the glass transition and the associated correlation are intrinsically related to thermodynamic anomalies accompanying the thermodynamic transition. 

As described above, the density $\rho$ is regarded as an important control variable in the glass transition. Note that the two processes of decreasing $T$ and increasing $\rho$ are generally related to each other \cite{density1,density2,density3}; intuitively, a decrease in the temperature reduces the overlap between the particles, which effectively increases the density. In (fragile) glass forming liquids, an increase in the density leads to strong steric hindrances for particle motions, based on which some theoretical models, such as the free-volume theory \cite{Cohen-Turnbull,Turnbull-Cohen} and the mode coupling theory (MCT) \cite{GotzeB}, have been proposed thus far. However, these models describe the slowing-down behavior as particle-scale phenomena, namely, without a strong concept of cooperativity or growing length scales \cite{MCT_comment1}. Certainly, the two-body density correlator hardly shows any anomalous or long range features despite the vast changes in the dynamic properties. This almost invariant property of the correlation of density fluctuations during the vitrification is in contrast to the increasing cooperativity, which may be one reason why, in recent literature, still-unknown thermodynamic anomalies and the associated growing static structures (if any) have been invoked for the origin of the growing cooperative length scale. 

However, some recent simulations have demonstrated an intimate link between the local density and the local dynamic properties. (i) The particle mobility is higher in lower density regions \cite{Dunleavy_Wiesner_Yamamoto_Royall,Li_Zhu_Sun}, indicating that small spatial variations of the local density are related to the dynamic heterogeneity (see also Appendix A). (ii) The present author found that, in supercooled states, there is a hydrodynamic correlation length $r_{\rm h}$, which is comparable to the dynamic heterogeneity size, and the density fluctuations slowly relax via length scale dependent diffusion \cite{FurukawaG3,FurukawaG4,FurukawaG5}: the relaxation of larger scale ($\gtrsim r_{\rm h}$) fluctuations exhibits diffusive decay, where $r_{\rm h}$ can be regarded as a unit size. On the other hand, smaller scale ($\lesssim r_{\rm h}$) fluctuations are subordinate to the collective dynamics for the duration of the structural relaxation. These observations (i) and (ii) prompt us to try a different approach based on non-singular density fluctuations. For this purpose, the following fact should be a key clue: in a fragile glass-forming liquid near the glass transition point, a small change in the {\it macroscopic} average density determines the {\it macroscopic} glass transition. Concomitantly, the density itself fluctuates in space. Based on this fact, one may imagine that even a slightly higher {\it local} density should make the {\it local} steric constraints more severe and thus determine the {\it local} glassy nature. Based on this perspective, we propose a simple model for describing the growing length scale accompanying the glass transition with a concept of clusters and without introducing any thermodynamic anomalies; namely, the local glassy nature may be simply controlled by the local density {\it on average}, which eventually determines the glass transition and the associated correlation \cite{polymer}.  Before proceeding, we note that in this study, the considered system is supposed to be a fragile glass-former. In strong glass-formers, the dynamics are less cooperative \cite{FurukawaG5,Vogel-Glotzer,Coslovich-Pastore,Kim-Saito}, and the role of density fluctuations in the relaxation mechanism appears to be different from that in fragile glass-formers \cite{FurukawaG5}. 

\section{Model and Analysis}
Some details of the key assumptions for our model are as follows: 

(1) {\it Cluster formation induced by local densification}: 
We assume that in a higher-density subsystem, in which the average density is above a certain threshold value, $\rho_{\rm c}$, particle rearrangements are strongly obstructed owing to stronger steric hindrances (or constraints); that is, the thermodynamic force cannot promote relaxation, and independent particle-activation is prohibited. Henceforth, such a subsystem is called a (glassy) cluster. More specifically, the steric constraints are assumed to be characterized by the subsystem-density, and once a cluster is formed, density fluctuations "inside" the cluster are transiently frozen for a sufficiently long time period, with the exception of small thermal vibrations.

(2) {\it{ No thermodynamic anomaly in density fluctuations}}: it is well known that even in deeply supercooled states, density fluctuations hardly show thermodynamic anomalies. Here, it is reasonable to assume that density fluctuations simply obey Gaussian statistics. 

Although other assumptions will be introduced in the following analysis, only these two assumptions are essential in constructing a model for the growing length scale.    

Let us consider the situation in which the glass transition point is approached by compression (increasing the macroscopic average density $\rho$ to $\rho_{\rm g}$) at a fixed temperature $T$. We then consider a subsystem with linear dimension $\ell$ and volume $V_\ell(=\ell^d)$, where $d$ is the spatial dimension. The density, $\rho_\ell$, averaged over the subsystem is given by 
\begin{eqnarray}
\rho_{\ell} = \rho + \dfrac{1}{V_\ell} \int_{V_\ell} d\mbox{\boldmath$r$} \delta \rho(\mbox{\boldmath$r$} ),
\end{eqnarray}
where $\delta \rho(\mbox{\boldmath$r$})$ is the (local) density fluctuation at position $\mbox{\boldmath$r$}$ from the average $\rho$. Because we now assume that $\delta \rho(\mbox{\boldmath$r$})$ obeys Gaussian statistics in thermal equilibrium, a fluctuation of $\rho_\ell$ is described as follows: 
\begin{eqnarray}
{K}\bigg\langle \biggl(\dfrac{ \rho_{\ell}-\rho}{\rho}\biggr)^2 \biggr\rangle\ell^d\cong T, \label{EQP1} 
\end{eqnarray} 
where the temperature $T$ is measured in units of the Boltzmann constant and $\langle \cdots \rangle$ denotes the ensemble average. Here, $K$ is the bulk modulus (inverse of the compressibility).
In deeply supercooled states, it should be appropriate to consider fluctuations in inherent states, for which $K$ should be replaced by the modulus, ${\bar K}$, of the inherent states. However, for this qualitative study, the difference between $K$ and ${\bar K}$ does not matter. In the following argument, taking only the fluctuation contribution to the leading order, we ignore the effect of density fluctuations on $K$. From Eq. (\ref{EQP1}), we obtain 
\begin{eqnarray}
\bigg\langle \biggl(\dfrac{ \rho_{\ell}-\rho}{\rho}\biggr)^2 \bigg\rangle \cong \biggl(\dfrac{a}{\ell}\biggr)^d , \label{EQP2} 
\end{eqnarray}
where $a= (T/K)^{1/d}$ is the microscopic length scale. At $ \langle ({ \rho_\ell-\rho})^2 \rangle \cong ({ \rho_{\rm c}-\rho})^2$, we can find a significant population of subsystems for which $\rho_\ell$ exceeds the threshold value $\rho_{\rm c}$ (see Fig. 1 for schematic); then, the size of such subsystems, $\xi$, is given by 
\begin{eqnarray}
\xi = a \biggl(\dfrac{\rho}{\rho_{\rm c}-\rho}\biggr)^{2/d}. 
\label{size_rho} 
\end{eqnarray}  
This $\xi$ gives the characteristic size of the glassy clusters. For example, in supercooled Lennard-Jones (or similar model) liquids, $a$ is estimaed to be several 0.1s of the unit of the particle size. Thus, when $\rho/(\rho_{\rm c}-\rho)$ is $10-100$, $\xi$ is approximately $1-5$, which appears to be reasonable. We emphasize again that $\xi$ is not the static correlation length of the density fluctuations determined by the two-body correlator, but instead measures how long the steric constraints persisit in blocking the rearrangement motions. \cite{comment_xi}.

\begin{figure}[tb] 
\includegraphics[width=0.425\textwidth]{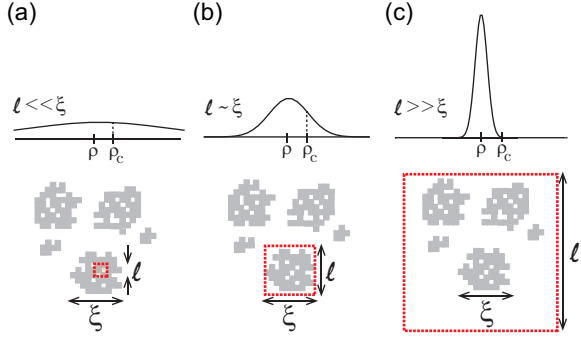} 
\caption{(Color online) Upper panels illustrate the probability distribution of $\rho_{\ell}$ for different $\ell$. The lower panels are the corresponding real space illustrations: in (a)/(c), the subsystem (red dashed box) is too small/large to correctly identify the glassy cluster (gray region). }
\label{Fig1}
\end{figure}

For smaller $\rho$, from Eq. (\ref{size_rho}), $\xi$ is smaller. There should be a minimum size of $\xi$ below which the concepts of density and cooperativity (or coherence) are no longer relevant. We set this lower bound value of $\xi$ to $\xi_{0}$, which may be comparable to the size (or diameter) of a group of nearest neighboring particles that form a cage ($\sim$ several particle sizes). In this context, the onset of cooperativity at $\rho\cong \rho_{0}$ is identified as an emergence of glassy clusters of the size $\xi_0$ with a significant volume fraction.

Next, let us refine the estimate of $\xi$ by considering the cluster-size distribution. It is a convenient simplification, without loss of generality, to consider a discrete sequence of sizes (volumes) $\xi_{(n)}$ ($\xi_{(n)}^d$), $n=1,2,3, \cdots$, of the glassy clusters as follows. In this discretization scheme, $\xi_{(1)}$ is defined as 
\begin{eqnarray}
\xi_{(1)}= \Lambda \xi, \label{1st_cluster}
\end{eqnarray}
where $\Lambda$ is a constant of order unity and controls the fineness of the discretization (finer for larger $\Lambda$).  Although the clusters are randomly generated, in the following analysis, we will identify the clusters from larger ones. According to Eqs. (\ref{EQP2}) and (\ref{size_rho}), for a subsystem with linear dimension $\xi_{(1)}$, we may define the probability distribution of $\rho_{\xi_{(1)}}$ as 
\begin{eqnarray}
P(\rho_{\xi_{(1)}})=\sqrt{\dfrac{\Lambda^d}{2\pi (\rho_{\rm c}-\rho)^2}} \exp\biggl[-\dfrac{\Lambda^d}{2}\biggl(\dfrac{\delta\rho_{\xi_{(1)}}}{\rho_{\rm c}-\rho}\biggr)^2\biggr], 
\end{eqnarray}
where $\delta\rho_{\xi_{(1)}}=\rho_{\xi_{(1)}}-\rho$. 
The probability that a given subsystem of size $\xi_{(1)}$ is a glassy cluster ($\rho_{\xi_{(1)}}>\rho_{\rm c}$) is given by 
\begin{eqnarray}
\phi_{(1)}&=&\phi= \int_{\rho_{\rm c}-\rho}^\infty d(\delta\rho_{\xi_{(1)}}) P(\rho_{\xi_{(1)}}) \nonumber \\
&=& \dfrac{1}{\sqrt{\pi}}\int_{x_0}^\infty dx \exp(-x^2),   
\label{volume_fraction}
\end{eqnarray}
where $x_0=\sqrt{\Lambda^d/2}$. 
Therefore, we can find $1/(\phi\xi^3_{(1)})$ clusters in a unit volume. The density averaged over the cluster regions, $\rho_{+}^{(1)}$, is  
\begin{eqnarray}
\rho_{+}^{(1)}&=&\rho+\dfrac{1}{\phi} \int_{\rho_{\rm c}-\rho}^\infty d(\delta\rho_{\xi_{(1)}}) \delta\rho_{\xi_{(1)}} P(\rho_{\xi_{(1)}}) \nonumber \\ 
&=& \rho + \lambda (\rho_{\rm c}-\rho),  
\end{eqnarray}
where $\lambda= \frac{1}{2\sqrt{\pi} \phi x_0}\exp(-x_0^2)$. On the other hand, in the remaining space, ${\mathcal V}_{-}^{(1)}$ (see Fig. 2(a) for schematic), the average density is given by 
\begin{eqnarray}
\rho_{-}^{(1)}= 
\dfrac{1}{1-\phi}\rho- \dfrac{\phi}{1-\phi}\rho_+^{(1)} = \rho_{\rm c} - \nu (\rho_{\rm c}-\rho), \label{except_1st}
\end{eqnarray}
where $\nu =(1-\phi+\lambda\phi)/({1-\phi})$. 
\begin{figure}[tb] 
\includegraphics[width=0.38\textwidth]{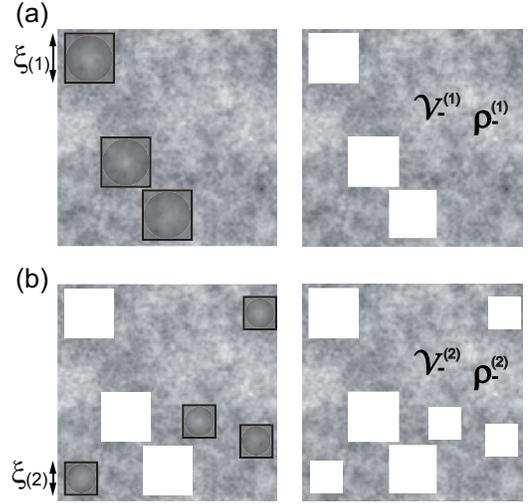} 
\caption{Schematic of the cluster distribution: (a) Left panel: Clusters of size of $\xi_{(1)}$ are shown by the dark-gray regions, where the average density $\rho_{\xi_{(1)}}$ is larger than $\rho_{\rm c}$. Right panel: In the remaining space, ${\mathcal V}_{-}^{(1)}$, the average density is $\rho_{-}^{(1)} = \rho_{\rm c} -\nu (\rho_{\rm c}-\rho)$. (b) Left panel: For $\rho_{-}^{(1)}>\rho_0$, we can find smaller clusters of the size $\xi_{(2)}=\nu^{-2/d}\xi_{(1)}(>\xi_0)$ in ${\mathcal V}_{-}^{(1)}$, which are represented by the dark-gray regions. Right panel: ${\mathcal V}_{-}^{(2)}$ is defined to be the space exterior to the clusters of the first and second steps. In ${\mathcal V}_{-}^{(2)}$, the average density is given by $\rho_{-}^{(2)} \cong \rho_{\rm c}-\nu^2(\rho_{\rm c}-\rho)$. If $\rho^{(2)}_{-} > \rho_0$, in ${\mathcal V}_{-}^{(2)}$, further smaller clusters can be found.}
\label{Fig2}
\end{figure}
Note that although the values of $\phi$, $\lambda$, and $\nu$ depend on the details of the discretization scheme employed here (for example $\phi\cong 0.16$, $\lambda\cong 1.53$, and $\nu\cong1.29$ for $\Lambda=1$), the final conclusion does not depend on these values. For $\rho\cong \rho_{0}$, $\rho_-^{(1)}$ is significantly smaller than $\rho_{0}$; thus, in $\mathcal V_{-}^{(1)}$, almost uncorrelated motions of particles should occur as in normal liquid states. However, for sufficiently large $\rho$, this $\rho_{-}^{(1)}$ can be significantly larger than $\rho_0$. In such a case, in ${\mathcal V}_{-}^{(1)}$, we can find smaller clusters of size $\xi_{(2)}(<\xi_{(1)}$) with a density larger than $\rho_{\rm c}$. 
Similarly to Eq. (\ref{1st_cluster}), $\xi_{(2)}$ may be defined as     
\begin{eqnarray}
\xi_{(2)} = a \Lambda \biggl(\dfrac{\rho_{-}^{(1)}}{\rho_{\rm c}-\rho_{-}^{(1)}}\biggr)^{2/d} \cong \nu^{-2/d}\xi_{(1)}. 
\label{size_rho2} 
\end{eqnarray}
In Eq. (\ref{size_rho2}) and subsequently, we keep only the leading-order term in $(\rho_{\rm c}-\rho)$. In ${\mathcal V}_{-}^{(1)}$, a density averaged over a subsystem with the linear size $\xi_{(2)}$, $\rho_{\xi_{(2)}}$, has the following distribution: 
\begin{eqnarray}
P(\rho_{\xi_{(2)}}) = \sqrt{\dfrac{\Lambda^d}{{2\pi} (\rho_{\rm c}-\rho_{-}^{(1)})^2}} \exp\biggl[-\dfrac{\Lambda^d}{2}\biggl(\dfrac{\delta\rho_{\xi_{(2)}}}{\rho_{\rm c}-\rho_{-}^{(1)}}\biggr)^2\biggr], \nonumber \\
\end{eqnarray}
where $\delta\rho_{\xi_{(2)}} = \rho_{\xi_{(2)}}-\rho_{-}^{(1)}$. Then, we can find $1/(\phi\xi_{(2)}^3)$ clusters of size $\xi_{(2)}$ per unit volume in ${\mathcal V}_{-}^{(1)}$. Similar to Eq. (\ref{except_1st}), in the space other than that occupied by the clusters of the first and second steps, ${\mathcal V}_{-}^{(2)}$ (see Fig. 2(b) for schematic), the average density is given by 
\begin{eqnarray}
\rho_{-}^{(2)} &\cong&  \rho_{\rm c} - \nu (\rho_{\rm c}-\rho_{-}^{(1)}) =  \rho_{\rm c}-\nu^2(\rho_{\rm c}-\rho), \label{except_2nd}
\label{rho2} 
\end{eqnarray} 
If $\rho^{(2)}_{-} < \rho_0$, in ${\mathcal V}_{-}^{(2)}$, the rearrangement dynamics proceed by almost independent particle motions, as in normal liquid states. However, if $\rho^{(2)}_{-}$ is significantly larger than $\rho_0$, further smaller clusters can be found. In this discretization scheme, at the $n$-th step, the size of the clusters, $\xi_{(n)}$, and the average density in ${\mathcal V}_{-}^{(n)}$, which is defined to be the space exterior to the clusters from the first to the $n$-th step, $\rho_{-}^{(n)}$, can be described by \cite{comment_nth}
\begin{eqnarray}
\xi_{(n)}\cong \nu^{-2(n-1)/d}\xi_{(1)},
\end{eqnarray} 
and
\begin{eqnarray}
\rho_{-}^{(n)} \cong  \rho_{\rm c}-\nu^n(\rho_{\rm c}-\rho) ~~ {\rm in}~~{\mathcal V}_{-}^{(n)},  
\label{size_rhon} 
\end{eqnarray} 
respectively. When $\rho^{(n)}_{-}\cong \rho_0 $ in ${\mathcal V}_{-}^{(n)}$, which leads to $\xi_{(n)}\sim\xi_0$, the particles move almost independently, as in normal liquid states, for which we set $n=\mathcal N$: 
\begin{eqnarray}
{\mathcal N}= \dfrac{1}{\ln \nu}\ln \biggl(\dfrac{\rho_{\rm c}-\rho_0}{\rho_{\rm c}-\rho}\biggr). \label{number} 
\end{eqnarray}
The volume fraction of the $n$-th step clusters is $\phi(1-\phi)^{n-1}$. 
Thus, the volume fraction of the cluster region $\Phi$ is 
\begin{eqnarray}
\Phi&\cong& \sum_{n=1}^{\mathcal N}\phi(1-\phi)^{n-1}=1-(1-\phi)^{\mathcal N}, \nonumber \\ &=& 1- \biggl(\dfrac{\rho_{\rm c}-\rho}{\rho_{\rm c}-\rho_0}\biggr)^{\mu}, 
\label{cluster_fraction}
\end{eqnarray}
where $\mu=-[\ln(1-\phi)/\ln\nu]$. The average cluster size is 
\begin{eqnarray}
\bar\xi &\cong& \dfrac{1}{\Phi}\sum_{n=1}^{\mathcal N}\phi(1-\phi)^{n-1}\xi_{(n)} . 
\end{eqnarray} 
With increasing macroscopic average density $\rho$, $\mathcal N$ also becomes larger, and the contributions from the larger clusters are dominant, resulting in $\bar\xi\sim \xi$.

 Our model shows that the length scale diverges as $(\rho_{\rm c}-\rho)^{-2/d}$. The exponent describing this divergence is the same as the one predicted by several theoretical models \cite{Tanaka_review,Kirkpatrick_Thirumalai_Wolynes,Lubchenko_WolynesR,Tanaka-Kawasaki-Shintani-Watanabe} and indeed has been observed in some simulations \cite{Tanaka-Kawasaki-Shintani-Watanabe,Malins_Eggers_Tanaka_Royall,Mosayebi_DelGado_Ilg_Ottinger,Mosayebi_DelGado_Ilg_Ottinger2} and experiments \cite{Zhang_etal,Weingartner_Soklaski_Kelton_Nussinov}. However, the basic mechanism considered here is very different: our premise is that the {\it local} glassy nature is simply controlled by the {\it local} subsystem density on average; via compression, the characteristic size of the glassy clusters increases, whereas the static properties of the density fluctuations remain almost unchanged (see Fig. 3 for schematic). We infer that strong thermodynamic anomalies and their associated intrinsic long-range correlation found in spin-glasses and critical phenomena are absent even in deeply supercooled states. 
\begin{figure}[t] 
\includegraphics[width=0.4\textwidth]{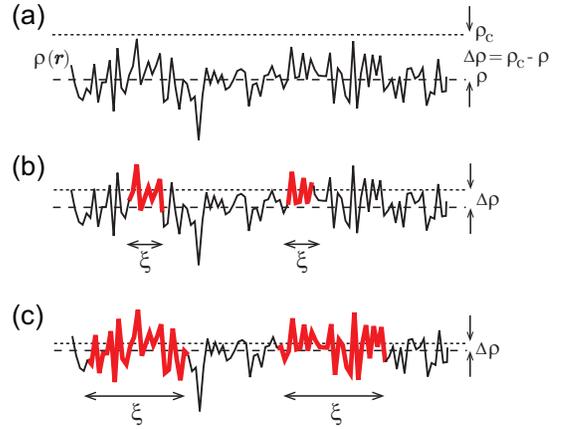} 
\caption{(Color online) Schematic of the growing length scale: the density $\rho(\mbox{\boldmath$r$})$ fluctuates around the average value $\rho$ (shown as dashed lines). The cluster regions, in which the average density exceeds the threshold value $\rho_{\rm c}$ (shown as dotted lines), are represented by thick red lines. With increasing $\rho$ (from (a) to (c)), $\xi$ increases, whereas the static properties of the density fluctuations remain almost unchanged. }
\label{Fig3}
\end{figure}

Here, we briefly consider the situation in which the glass transition point is approached by decreasing $T$ at a fixed $\rho$. As noted in Sec. I, the two processes, increasing $\rho$ and decreasing $T$, are generally related to each other \cite{density1,density2,density3}. Because weakening the thermal fluctuations effectively increases the density, the threshold density $\rho_{\rm c}$ is decreased as the temperature decreases toward $T_{\rm c}$; at $T=T_{\rm c}$, the average density $\rho$ is identical to $\rho_{\rm c}$. For $T\cong T_{\rm c}$, $\rho_{\rm c}(T)$ can be expanded as   
\begin{eqnarray}
\rho_{\rm c}(T)\cong \rho +\dfrac{\partial \rho_{\rm c}}{\partial T}\bigg|_{T=T_{\rm c}}(T-T_{\rm c})\cdots,   \label{close_Tc}
\end{eqnarray}
which is expected to hold generally near the glass transition point. 
Therefore, Eq. (\ref{size_rho}) can be rewritten as 
\begin{eqnarray}
\xi \sim a' \biggl(\dfrac{T}{T - T_{\rm c}}\biggr)^{-{2}/{d}},   \label{size_t} 
\end{eqnarray}
where $a'=a [(\partial \ln\rho_{\rm c}/\partial \ln T)|_{T=T_{\rm c}}]^{2/d}$ should again be the microscopic length scale. It is known that a large number of molecular glass-formers show the isomorph scaling \cite{Dyre1,Dyre2}, $\rho^\Gamma/T={\rm const.}$, where $\Gamma$ is a material dependent parameter. Through this scaling relation, the state $(\rho_{\rm c},T)$ can be mapped onto the state $(\rho,T_{\rm c})$, and then, the temperature dependence of the threshold density $\rho_c$ should be of the form 
\begin{eqnarray}
\rho_c= \rho\biggl(\dfrac{T}{T_{\rm c}}\biggr)^{1/\Gamma}, \label{isomorph}
\end{eqnarray}
which is reduced to Eq. (\ref{close_Tc}) at $T\cong T_{\rm c}$ with $(\partial \rho/\partial T)|_{T=T_{\rm c}}=\rho/\Gamma T_{\rm c}$. 

\section{Discussion on the dynamics}
Finally, we provide a sketch of the supercooled liquid dynamics based on the clusters. At $\rho\cong\rho_0$, where the volume fraction of the clusters is small, the structural relaxation should mainly proceed in the ``normal liquid'' region, where fast and almost independent particle motions are allowed to occur. However, at a higher density, for which the volume fraction of the clusters occupies a larger space, structural relaxation involving the cluster dynamics should be more dominant: Thus far, in this study, we have supposed that, in deeply supercooled states, ``inside'' the cluster, independent activation at the particle-scale is highly suppressed owing to steric hindrances (except for the thermal rattling motions of the particles), which implicitly assumes that the structural rearrangements should occur cooperatively and thus that the cluster lifetime is comparable to or is longer than the structural relaxation time $\tau_\alpha$ \cite{comment_cluster_life}.  

Let us consider a deeply supercooled state with a high volume fraction of clusters of typical size $\xi$. The average configuration of the clusters may remain unchanged for small thermal activation, whereas for sufficiently large thermal activation, the cluster configuration may be unstable and then undergo rearrangement as a cooperative event. Such a rearrangement may control the structural relaxation. Supposing that a cluster is (transiently) immersed in an effective elastic medium, the restoring force acting on the cluster is approximately $G\xi\gamma$, where $\gamma$ is the displacement amplitude and $G\xi$ is a force constant with $G$ being the shear elastic modulus of the bulk system. Thus, the elastic deformation energy of the medium ($E_{\rm el}^{\rm (m)}$) is estimated to be $E_{\rm el}^{\rm (m)}\sim G\xi\gamma^2$. On the other hand, the elastic energy due to the cluster deformation ($E_{\rm el}^{\rm (c)}$) is estimated to be $E_{\rm el}^{\rm (c)}\sim G\xi^3 \times (\gamma/\xi)^2=G\xi\gamma^2$, where $\gamma/\xi$ is the typical strain in the deformed cluster. Thus, $E_{\rm el}^{\rm (m)}$ and $E_{\rm el}^{\rm (c)}$ share the same order of magnitude. For the (tagged) cluster, the restoring energy due to elastic deformation involving the surrounding "medium" can be given as $E_{\rm el} \sim G\xi \gamma^2$. 
When the thermal activation is sufficiently large so that $\gamma$ is comparable to $\xi$ ($\gamma\cong c\xi$ with $c$ being a small factor), a transition from one cluster configuration to another may occur. For this significant rearrangement event, the activation energy is simply estimated as 
\begin{eqnarray}
\Delta E_{\rm el}\sim G \xi^3 \sim T \biggl(\dfrac{\xi}{\xi_0}\biggr)^3.  \label{activation_energy}
\end{eqnarray} 
This argument is similar to the one for the model describing the activation energy of a particle jump \cite{Hall_Wolyness,Buchenau_Zorn,DyreR}. However, we emphasize again that the individual particle activation should be suppressed in a deeply supercooled state due to the severe topological restrictions. Combining Eq. (\ref{activation_energy}) with Eq. (\ref{size_rho}), we expect the relaxation time to diverge as 
\begin{eqnarray}
\tau_\alpha(\rho) \sim e^{ -\kappa [{\rho}/({\rho_{\rm c}-\rho})]^2}. \label{time_rho}
\end{eqnarray}
Similarly, in the case in which the temperature is decreased at fixed density, we also have 
\begin{eqnarray}
\tau_\alpha(T) \sim e^{ -\kappa' [{T}/({T-T_{\rm c}})]^2}. \label{time_T}
\end{eqnarray}
Here, $\kappa$ and $\kappa'$ are numerical constants. While Eqs. (\ref{time_rho}) and (\ref{time_T}) exhibit stronger divergences than the standard Vogel-Fulcher-Tamman form, some experiments report that the form of Eq. (\ref{time_rho}) is preferred \cite{Brambilla_etal}. According to Eq. (\ref{activation_energy}), when $\xi$ is $3-4$ times larger than $\xi_0$, the activation energy is $20-30$ times larger than that at the crossover state, resulting in an increase in $\tau_\alpha$ by more than $10$ orders of magnitude. 

We note that in the literature \cite{Goldstein,DyreR,Dyre_Christensen_Olsen,Anael1,Anael2}, it was argued that the relaxation process in supercooled liquids consists of the elasticity-driven consecutive transition between inherent states and that the accumulation of many such transition events will manifest as hydrodynamic relaxation, which inspired the current argument \cite{comment_difference}.

\section{Concluding Remarks}

In this paper, we have constructed a phenomenological model for describing a growing length scale accompanying the vitrification: we have assumed that in a subsystem whose density is above a certain threshold value, $\rho_{\rm c}$, the particle rearrangements are highly suppressed and the dynamical coherence is maintained for a certain long time-period owing to steric hindrances (or restrictions). With this assumption and without invoking thermodynamic anomalies, we have predicted that upon compression (increasing the average density $\rho$) at a fixed temperature $T$ in supercooled states, the characteristic length of the clusters, $\xi$, diverges as $\xi\sim(\rho_{\rm c}-\rho)^{-2/d}$. Additionally, with decreasing $T$ at fixed $\rho$, the length scale diverges as $\xi\sim(T-T_{\rm c})^{-2/d}$, for which, at $T=T_{\rm c}$, $\rho$ is identical to $\rho_{\rm c}$. The exponent describing the diverging length scale is the same as the one predicted by certain previous theoretical models \cite{Kirkpatrick_Thirumalai_Wolynes,Lubchenko_WolynesR,Tanaka-Kawasaki-Shintani-Watanabe}, but the basic mechanism for the divergence is different.  

Several other theoretical models assume that thermodynamic anomalies are not involved in the glass transition. Here, we make some remarks regarding two such models, namely kinetically constrained models (KCMs) \cite{Ritort-SollichR,Garrahan-ChandlerR} and mode coupling theory (MCT) \cite{GotzeB}: 

Kinetically constrained models (KCMs) are known to reproduce many aspects of supercooled-liquid dynamics. Most notably, KCMs show that heterogeneity and cooperativity in the dynamics can be of a purely dynamical origin; in other words, thermodynamics may not play a role in the main characteristics of supercooled-liquid dynamics. At this stage, an exact relationship between our model and KCMs is not clear. In this study, while we have argued that singular dynamics observed in deeply supercooled liquids are linked with non-singular equilibrium density fluctuations, the statistics of which are determined by thermodynamics, we have not ascribed the origin of such singular dynamics to a purely kinetic effect. Furthermore, KCMs basically do not suggest any singularity of the relaxation time at finite temperature \cite{comment_KCMs}, whereas here the finite temperature singularity is considered by the limiting density. This difference is significant. 

 Mode coupling theory (MCT) \cite{GotzeB} also supposes that {singular} dynamics of supercooled liquids are directly related to {non-singular} density fluctuations. However, the standard MCT does not include any concept of growing length scale or heterogeneity. In MCT, the mode corresponding to the length scale of the static density correlation dominates the dynamics. For systems such as normal simple liquids and critical fluids, where the static correlation length is identified with the relevant length scale for the dynamics, the MCT scheme provides a very good approximation for calculating transport coefficients and their length scale dependences. However, this is not the case for supercooled liquids, where with the increasing degree of supercooling, the static correlation length ($\sim$ the particle size) increasingly deviates from the dynamic one. As argued in Sec. II, rather than the correlation length of density fluctuations, the persistence length of steric constrains, assumed to be determined by the subsystem-density, should be important for describing the local vitrification. Such, so to speak, cooperative caging or jamming cannot be described by the present MCT. In addition, we note that it is still not known whether the generalized hydrodynamic equations employed to construct the present MCT are sufficient for the description of the supercooled liquid dynamics.

Before closing, we note the following points. 

(i) In the very recent study by the present author, a simple model for shear-thinning in a high-density glassy liquid was proposed \cite{FurukawaG6}: in a shear flow, due to the asymmetric shear flow effect on particles, the effective density is reduced. Because $\tau_\alpha$ depends strongly on the density near the glass transition point, even a very small reduction in the effective density significantly accelerates the structural relaxation. In the context of the present study, this shear-induced reduction of the effective density would be accompanied with a decrease in the cluster size and thus drive the system away from the glass transition point. In some simulation studies of supercooled liquids \cite{Yamamoto_Onuki,Mizuno_Yamamoto}, it was found that the dynamic heterogeneity sizes are decreased when shear-thinning occurs, which may support our argument. 

(ii) In this paper, each glassy cluster has been assumed to be almost independent. However, it may be possible that the clusters percolate to form a ramified network structure at the threshold value of the cluster volume-fraction, $\Phi_{\rm p}$: for $\Phi\gtrsim \Phi_{\rm p}$, the clusters are not closely packed, and thus, the formed network structure should not be rigid enough to prevent macroscopic relaxation; that is, the clusters should still be almost independent. However, for $\Phi\gg \Phi_{\rm p}$, before $\rho$ reaches $\rho_{\rm c}$, the developing  network structure may be sufficiently thick to freeze the macroscopic dynamics; in such a situation, the growing length scale is not given by $\xi$ but may be characterized by, for example, the stress correlation associated with the cluster-percolation. 

(iii)
Some authors have argued that locally favored structures observed in some kind of glass formers indicate a thermodynamic competition between different states \cite{Tarjus,Paddy}. However, because such structures should be sensitive to the "local" packing fraction, We infer that locally favored structures may simply reflect the fluctuations of the subsystem density. 

(iv)
It is often stated that, at the hypothetical Kauzmann temperature (or the corresponding density), a liquid is supposed to be at the {\it ideal glass transition point} characterized by a single thermodynamic configuration, that is, a macroscopically unique equilibrium state. However, as argued in this paper, our interpretations of the vitrification and the associated singular behavior are different from those based on thermodynamic anomaly. In this paper, we have assumed that when the subsystem-density is above the threshold value, the thermal activation and thermodynamic force cannot promote the relaxation; consequently, the subsystem is transiently trapped in a frozen state \cite{MCT_comment2}. In this context, the threshold density $\rho_c$ (or the corresponding temperature $T_c$) has been tacitly assumed to be lower (higher) than that at the Kauzmann point. Furthermore, regarding the macroscopic glass transition point, to block the macroscopic relaxation, a space-spanning (not space-filling, which may be excessive) steric constraint should be formed. If such a macroscopic constraint is realized as a metastable state, then a significant number of configurations can be considered even at the glass transition point. In this view, the glass transition density (temperature) is also lower (higher) than that at the Kauzmann point. However, at this stage, we cannot comment on the precise physical meaning of $\rho_c$ and whether the local and global singular points coincide or are close to each other; in this study, the existence of $\rho_c$ is simply presupposed. Further theoretical and numerical investigations on these remaining fundamental and difficult problems are the subject of another study. 

We will examine these speculations in future work. 

The author thanks Professors Hajime Tanaka, Jeppe C. Dyre, Hajime Yoshino, and Atsushi Ikeda for useful comments. This work was supported by KAKENHI (Grant No. 26103507, No. 25000002, and No. 26400425) and the JSPS Core-to-Core Program ``International research network for non-equilibrium dynamics of soft matter''.  

\appendix
\section{The intimate link between local density and local (im)mobility}

\subsubsection{Model details} 
Several results have been reported  \cite{Dunleavy_Wiesner_Yamamoto_Royall,Li_Zhu_Sun} on the link between local particle (im)mobility and local density \cite{Dunleavy_Wiesner_Yamamoto_Royall,Li_Zhu_Sun}. In this Appendix, we provide clearer evidence for such a link based on the three-dimensional simulation results of a model glass-forming liquid, namely, Bernu-Hiwatari-Hansen (BHH) soft-sphere model \cite{Bernu-Hiwatari-Hansen}. This model has been thoroughly studied by many authors \cite{Bernu-Hiwatari-Hansen,Bernu-Hansen-Hiwatari-Pastore,Roux-Barrat-Hansen,Yamamoto-Onuki,FurukawaV1,FurukawaV2}. The BHH model is a binary mixture of small (species $1$) and large (species $2$) particles interacting via the following soft-core potentials $U_{ab}(r)=\epsilon ({s_{ab}}/{r})^{12}$, where $a,b= 1,2$, $s_{ab}=(s_{a}+s_{b})/2$, $s_{a}$ is the particle size, and $r$ is the distance between two particles. The mass and size ratios are $m_{2}/m_{1}=2$ and $s_{2}/s_{1}=1.2$, respectively. The units for the length and time are $s_1$ and $({m_{1}s_{1}^{2}/\epsilon})^{1/2}$, respectively. The temperature $T$ is measured in units of $\epsilon/k_{B}$, where $k_{B}$ is the Boltzmann constant. The total number of particles is $N=N_{1}+N_{2}$ and $N_1/N_2=1$, with $N_{a}$ being the number of particles of species $a$. The fixed particle number density of the system is $n_0=N/V =0.8$. Here, we set $N$=40000 (or 320000) and $V^{1/3}=36.84$ (or 73.86). In the preset binary system, the effective one-component density at time $t$ is given by \cite{Hansen_McdonaldB} $\rho(\mbox{\boldmath$r$},t)=s_1^3n_1(\mbox{\boldmath$r$},t)+s_2^3n_2(\mbox{\boldmath$r$},t)$, where $n_1$ and $n_2$ are the number densities of species 1 and 2, respectively.
The density in a subsystem $V_\ell=\ell^3$ with linear dimension $\ell$ is defined as 
\begin{eqnarray}
\rho_{\ell}(t) =  \dfrac{1}{V_\ell}\int_{V_{\ell}} d \mbox{\boldmath$r$} \rho(\mbox{\boldmath$r$},t). \label{subsystem_density} 
\end{eqnarray} 

\subsubsection{immobility determined by overlapping}
The immobility of the $i$-th particle is defined as 
\begin{eqnarray}
{q}_{i}(\Delta t)=\theta(w-|{\mbox{\boldmath$r$}}_i(t_0+\Delta t)-{\mbox{\boldmath$r$}}_i(t_0)|), \label{overlap}
\end{eqnarray}
where $\theta$ is a step function, and $|{\mbox{\boldmath$r$}}_i(t_0+\Delta t)-{\mbox{\boldmath$r$}}_i(t_0)|$ is the absolute value of the displacement of the $i$-th particle over time $\Delta t$. We set $w=0.25$, which is comparable to the plateau value of the root of the mean square displacement of the constituent particles. Therefore, ${q}_{i}(t_0;\Delta t)=1$ indicates that the $i$-th particle at time $t_0$ and $t_0+\Delta t$ are almost overlapped; this particle is referred to as an immobile particle for the time interval $[t_0,t_0+\Delta t]$. Note that in the literature, instead of the immobility, the mobility, ($1-q_{i}$), is usually measured \cite{Berchier-BiroliR}; nevertheless, there is no essential difference in the observations. In Eq. (\ref{overlap}), to reduce the thermal vibration effects, the short-time averaged particle position is introduced, 
\begin{eqnarray}
\bar{\mbox{\boldmath$r$}}_i(t_0) = \dfrac{1}{\delta t} \int_{t_0}^{t_0+\delta t} dt' {\mbox{\boldmath$r$}}_i(t'), \label{time_average_position}
\end{eqnarray}
and then the immobility is redefined as
\begin{eqnarray}
\hat{q}_{i}(\Delta t)=\theta(w-|\bar{\mbox{\boldmath$r$}}_i(t_0+\Delta t)-\bar{\mbox{\boldmath$r$}}_i(t_0)|).
\end{eqnarray}
In the following analysis, $\Delta t$ and $\delta t$ are chosen to be comparable to the $\alpha$-relaxation and the initial decay times of the autocorrelation of the macroscopic shear stress, respectively, as shown in Fig. 4. Note here that in supercooled states longer-term measurements of the displacements are less sensitive to whether or not time averaging of particle position, Eq. (\ref{time_average_position}), is performed.  
\begin{figure}[htb] 
\includegraphics[width=0.37\textwidth]{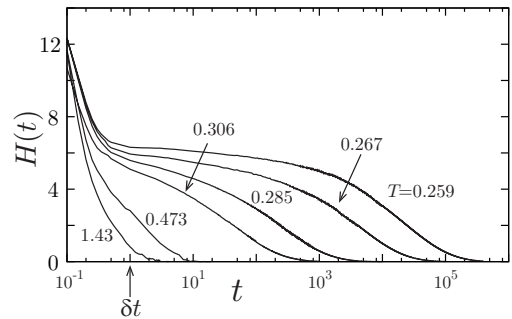}
\caption{
The autocorrelation of the macroscopic shear stress, $H(t)=(1/VT)\langle \sigma_{xy}(t)\sigma_{xy}(0)\rangle$, for several temperatures. Here, $\sigma_{xy}$ is the microscopic expression of the shear stress \cite{Hansen_McdonaldB}.  
The value of $\delta t$ is indicated by the arrow. In the present analysis, the $\alpha$ relaxation time $\tau_\alpha$ is defined as the relaxation time of $H(t)$.}
\label{FigS1}
\end{figure}

The immobility field is defined as 
\begin{eqnarray}
{\hat Q}(\mbox{\boldmath$r$};\Delta t)=\sum_{i} {\hat q}_{i}(t_0;\Delta t)\delta(\mbox{\boldmath$r$}-\bar{\mbox{\boldmath$r$}}_i(t_0)),  
\end{eqnarray} 
In Fig. 5, we plot the structure factor of the immobility field, $S_{\hat Q}(k,\Delta t)=(1/N)\langle|{\hat Q}_{\mbox{\boldmath$k$}}(t_0;\Delta t)|^2\rangle$, at $\Delta t=\tau_\alpha$ for three different temperatures, where ${\hat Q}_{\mbox{\boldmath$k$}}(\Delta t)$ is the Fourier transform of ${\hat Q}(\mbox{\boldmath$r$};\Delta t)$. This $S_{\hat Q}(k,\tau_\alpha)$ measures the spatial correlation of the particle (im)mobility.  
The low-$k$ behavior of $S_{\hat Q}(k,\tau_\alpha)$ can be fit to the following empirical function: 
\begin{eqnarray} 
S_{\hat Q}(k,\tau_\alpha) = \dfrac{S_0}{1+(k \xi_{\hat Q})^{x_{\hat Q}}}. 
\end{eqnarray}
In the literature, the exponent $x_{Q}$ is usually set to 2 by assuming the Ornstein-Zernike form of $S_{\hat Q}(k,\tau_\alpha)$. In this study, $x_{\hat Q}$ varies from 2.63 to 2.84 as the temperature is lowered from 0.306 to 0.259. Here, $\xi_Q$ is identical to the correlation length of the $\hat Q$ field, which increases as the temperature is lowered; $\xi_{\hat Q}=2.64$ ,3.41, and 3.65 for $T=$ 0.306, 0.267, and 0.259, respectively.   
\begin{figure}[htb] 
\includegraphics[width=0.325\textwidth]{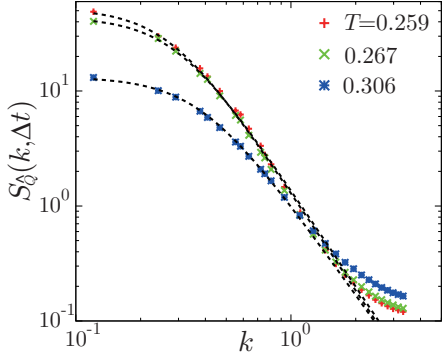}
\caption{
The structure factor of the immobility field, $S_{\hat Q}(k,\tau_\alpha)=(1/N)\langle|{\hat Q}_{\mbox{\boldmath$k$}}(t_0;\tau_\alpha)|^2\rangle$ for three different temperatures, $T=$ 0.259, 0.267, and 0.306. The black dashed curve represents the empirical fitting function, ${S_0}/{[1+(k \xi_{Q})^{x_{Q}}]}$. Here, $\xi_{\hat Q}=$ 3.65, 3.41, and 2.64 for $T=$ 0.259, 0.267, and 0.306, respectively. }
\label{FigS2}
\end{figure}

\subsubsection{The link between local immobility and local density}

\begin{figure}[t] 
\includegraphics[width=0.425\textwidth]{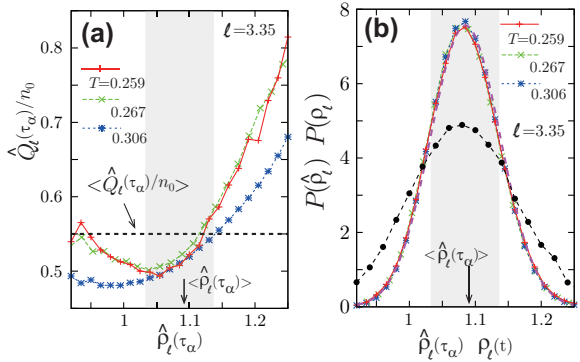}
\caption{
  (a) ${\hat\rho}_{\ell}(\tau_\alpha)$ vs. ${\hat Q}_{\ell}(\tau_\alpha)/n_0$ for $\ell=3.35$ at three different temperatures. $\langle {\hat Q}_{\ell}(\tau_\alpha) \rangle/n_0$ is shown by the black dashed line.
  (b) The probability distribution of ${\hat\rho}_{\ell}(\tau_\alpha)$, $P({\hat\rho}_{\ell})$, for $\ell=3.35$ at three different temperatures. The distribution function can be described by a Gaussian distribution represented by the solid purple curve; they are almost collapsed into a single curve. 
The probability distribution of the spontaneous density ${\rho}_{\ell}$, $P({\rho}_{\ell})$, for $\ell=3.35$ at $T=0.267$ is also shown (black circle and solid curve); this probability distribution is also described by a Gaussian distribution, but the variance is larger than the variance of the time-averaged density. In (a) and (b), the bright gray region represents $|\delta{\hat\rho}_{\ell}(\tau_\alpha)|< \sqrt{\langle \delta{\hat\rho}^2_{\ell}(\tau_\alpha)\rangle} $, where $\delta{\hat\rho}_{\ell}(\tau_\alpha)={\hat\rho}_{\ell}(\tau_\alpha)-\langle{\hat\rho}_{\ell}(\tau_\alpha)\rangle$. }
\label{FigS3}
\end{figure}

Let us examine the link between local immobility and local density. For this aim, we define the average immobility in a subsystem $V_\ell(=\ell^3)$ as  
\begin{eqnarray}
{\hat Q}_{\ell}(\tau_\alpha) =  \dfrac{1}{V_\ell}\int_{V_{\ell}} d \mbox{\boldmath$r$} {\hat Q}(\mbox{\boldmath$r$};\tau_\alpha).  
\end{eqnarray} 
In Fig. 6(a), we plot the subsystem immobility as a function of the time-averaged subsystem density  
\begin{eqnarray}
{\hat\rho}_{\ell}(\tau_\alpha) =  \dfrac{1}{\tau_\alpha} \int_{t_0}^{t_0+\tau_\alpha} dt'\rho_\ell(t'). 
\end{eqnarray}
where $\rho_\ell(t)$ is given in Eq. (\ref{subsystem_density}). In Fig. 6, we set $\ell=3.35$ ($\cong \xi_{\hat Q}$ at $T=$ 0.267). It is evident that the particles are more immobile (${\hat Q}_{\ell}(\tau_\alpha)\gtrsim \langle{\hat Q}_{\ell}(\tau_\alpha)\rangle$) in denser regions ($\delta{\hat\rho}_{\ell}(\tau_\alpha)\gtrsim \sqrt{\langle \delta{\hat\rho}^2_{\ell}(\tau_\alpha)\rangle}$), where $\delta{\hat\rho}_{\ell}(\tau_\alpha)={\hat\rho}_{\ell}(\tau_\alpha)-\langle{\hat\rho}_{\ell}(\tau_\alpha)\rangle$. 
This tendency is weaker at $T=0.306$ than at the lower two temperatures; at $T=0.306$ from Fig. 4, the stress autocorrelation does not exhibit a clear plateau, and thus the system is not sufficiently supercooled. 
In Fig. 6(b), we show the probability distribution of ${\hat\rho}_{\ell}(\tau_\alpha)$ for different temperatures at $\ell=3.35$. The distribution function can be described by a Gaussian distribution and shows a very small temperature dependence. Because ${\hat\rho}_{\ell}(\tau_\alpha)$ is time-averaged over $\tau_\alpha$, the thermal vibration effects are excluded \cite{comment_inherent}, and the variance of the distribution is smaller than that for the spontaneous subsystem-density ${\rho}_{\ell}(t)$, as shown in Fig. 6(b).  

\begin{figure}[bht] 
\includegraphics[width=0.425\textwidth]{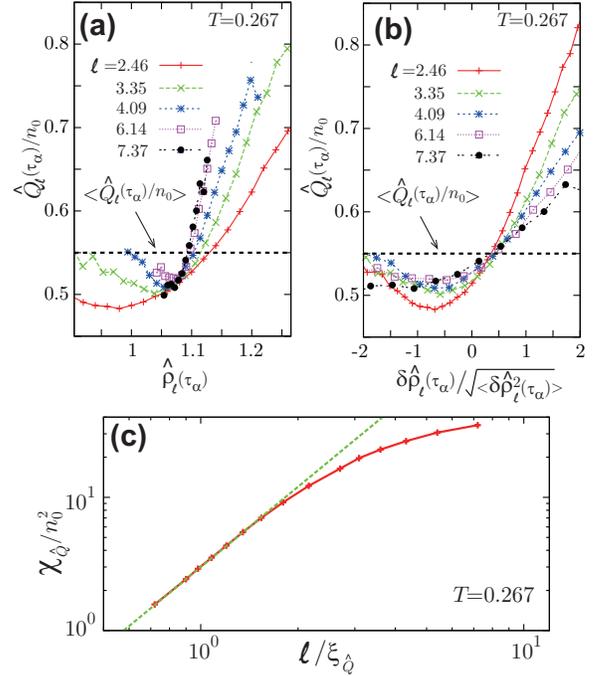}
\caption{
  (a) ${\hat Q}_{\ell}(\tau_\alpha)/n_0$ vs. ${\hat\rho}_{\ell}(\tau_\alpha)$ for various $\ell$ at $T=0.267$.
  (b) ${\hat Q}_{\ell}(\tau_\alpha)$ vs. $\delta{\hat\rho}_{\ell}(\tau_\alpha)/\sqrt{\langle \delta{\hat\rho}^2_{\ell}(\tau_\alpha)\rangle}$ for various $\ell$ at $T=0.267$. In (a) and (b), $\langle {\hat Q}_{\ell}(\tau_\alpha) \rangle/n_0$ is shown by the black dashed line. 
  (c) $\chi_{\hat Q}(\ell)= V_\ell[\langle {\hat Q}_{\ell}^2(\tau_\alpha)\rangle - \langle {\hat Q}_{\ell}(\tau_\alpha)\rangle^2]/n_0^2$ at $T=0.267$. The green dashed curve ($3(\ell/\xi_{\hat Q})^2$) is a fit to $\chi_{\hat Q}(\ell)/n_0^2$ for smaller $\ell(\lesssim 2\xi_{\hat Q})$. 
  }
\label{FigS4}
\end{figure}

Figures 7(a) and (b) show ${\hat Q}_{\ell}(\tau_\alpha)$ as a function of ${\hat\rho}_{\ell}(\tau_\alpha)$ and $\delta {\hat\rho}_{\ell}(\tau_\alpha)/\sqrt{\langle \delta {\hat\rho}_{\ell}^2(\tau_\alpha) \rangle}$, respectively, for different $\ell$ at $T=0.267$. For smaller $\ell$ each subsystem can be distinguished between mobile and immobile states, resulting in steeper ${\hat Q}_{\ell}(\tau_\alpha)$
in Fig. 7(b). Note that, over the range of $\ell$ investigated here, the probability distribution of ${\hat\rho}_{\ell}(\tau_\alpha)$ can be described by a Gaussian distribution.
In Fig. 7(c), the variance $\chi_{\hat Q}(\ell)= V_\ell[\langle {\hat Q}_{\ell}^2(\tau_\alpha)\rangle - \langle {\hat Q}_{\ell}(\tau_\alpha)\rangle^2]$ is plotted. The $\ell$ dependence of $\chi_{\hat Q}(\ell)$ is similar to that obtained in finite-size studies \cite{Karmakar_Dasgupta_Sastry1,Karmakar_Dasgupta_Sastry2}: $\chi_{\hat Q}$ increases, and then saturates for larger $\ell$; that is, for $\ell\gg \xi_{\hat Q}$ the subsystem has regions with a wider range of degrees of (im)mobbility with similar statistical properties.

\subsubsection{Short summary}
We have shown preliminary results indicating an intimate link between local density and local (im)mobility; the relationship between ${\hat Q}_\ell(\Delta t)$ and ${\hat \rho}_\ell(\Delta t)$ exhibits a tendency to show that denser regions are less mobile for the timescale of the structural relaxation. However, the present numerical results do not directly support the argument developed in the main text: The immobility ${\hat Q}_\ell(\Delta t)$ is determined by the total displacement for {\it a given time domain} $[t_0,t_0+\Delta t]$, and in each subsystem, immobile and mobile states are interchanged during the time period $\Delta t(\sim \tau_\alpha)$ with some probability. Therefore, the present measurement does not distinguish between "glassy" and "non-glassy" regions for {\it a given state at time} $t$. To examine the validity of our model, a different analysis using a specifically designed simulation setup is desirable, which will be the subject of a future study.


\begin{thebibliography}{70}
\bibitem{MCT_comment1}
There have been some efforts \cite{Biroli_Bouchaud_Miyazaki_Reichman,Berthier_Biroli_Bouchaud_Kob_Miyazaki_Reichman} to incorporate spatially inhomogeneous dynamics by extending the original MCT. However, the predicted growth is significantly weaker than those found in practical simulations and experiments.
\bibitem{polymer}
 Let us recall the slow dynamics of entangled polymers: entanglements are purely topological objects that are characterized by molecular weights and polymer concentrations but are not described through the free energy. The resultant topological constraints determine the slow structural relaxation and the dynamical correlation in an entangled polymer system \cite{deGennesB,Doi-EdwardsB,Brochard-deGennes}. This situation of entangled polymer dynamics reminds us that the slow dynamics is not always accompanied by thermodynamic anomalies. Actually, in the original concept of dynamic heterogeneity, it was considered that the observed correlation in the supercooled liquid dynamics does not have a thermodynamic origin. There is of course no direct relationship between these two phenomena, the supercooled liquid dynamics and the entangled polymer dynamics, but this apparent analogy tempted us to try the current approach.
\bibitem{comment_xi}
At smaller length scales, the density fluctuations are rather messy: Because the correlation length of the density fluctuations remains almost unchanged, there may be spots or regions with lower packing inside the glassy cluster. However, the induced steric constraints, ``on the whole'', are supposed to be enough to block the rearrangement motions in the cluster.
\bibitem{comment_nth}
In $\mathcal V_{-}^{(n-1)}$, a density averaged over a subsystem with linear size $\xi_{(n)}$, $\rho_{\xi_{(n)}}$, has the following distribution: $P(\rho_{\xi_{(n)}}) = \frac{1}{\sqrt{2\pi} (\rho_{\rm c}-\rho_{-}^{(n-1)})} \exp[-\frac{1}{2}(\frac{\delta\rho_{\xi_{(n)}}}{\rho_{\rm c}-\rho_{-}^{(n-1)}})^2]$, where $\delta\rho_{\xi_{(n)}}=\rho_{\xi_{(n)}}-\rho_{-}^{(n-1)}$ and $\rho_{-}^{(n-1)}$ is the average density in $\mathcal V_{-}^{(n-1)}$.
\bibitem{comment_cluster_life}
 In (fragile) glass-formers, the generation and relaxation of density fluctuations occur very slowly via diffusion \cite{FurukawaG3,FurukawaG4,FurukawaG5}. In supercooled states with significant numbers of glassy clusters, there may be two possible mechanisms for the density relaxations. One is normal diffusion: a higher-density (frozen) state inside the cluster relaxes via particle-exchange through the cluster surface, which occurs when a sufficiently low density region reaches the cluster's neighborhood. The other is cooperative diffusion: relaxation of density fluctuations occurs as cooperative rearrangements of the clusters. Our previous simulations \cite{FurukawaG3,FurukawaG4,FurukawaG5} indicated that the latter scenario is more likely.
\bibitem{comment_difference}
The relaxation mechanism discussed here is not consistent with one previously proposed by the present author in Ref. \cite{FurukawaG4}, which is different from cooperative activation.
\bibitem{comment_KCMs}
In KCMs, it is demonstrated that, as the temperature is lowered, local kinetic constraints make the dynamics slower, but unless all of the sites (particles) are inactive as realized at $T=0$, the macroscopic relaxation certainly proceeds.
\bibitem{MCT_comment2}
In this study, we have assumed that when the subsystem-density is above the threshold value, the subsystem is transiently trapped in a (metastable) frozen state. In this sense, the meaning of $\rho_c$ in our model is similar to that of the MCT critical density, but the relationship between our model and MCT is not clear. It is known that the MCT critical density is significantly lower than the density at the actual singular point, which is evaluated from experimental or simulation data. In addition, as noted in Sec. I, MCT describes an even weaker growth of length scales. In the literature, these discrepancies have been ascribed to some insufficiencies of the MCT approximation. Future studies should be devoted to further investigating these points, which is beyond the scope of this paper.
\bibitem{comment_inherent}
The physical meaning of ${\hat\rho}_{\ell}(\tau_\alpha)$ should be similar to that of the local density in inherent states; that is, by the time average over the time period of $\tau_\alpha$, the mean local density is extracted while removing the thermal fluctuation effects. 
The relaxation time of density fluctuations on the length scale of $\ell$ investigated here is significantly longer than that of the shear stress (=$\tau_\alpha$ here) \cite{FurukawaV2}. This timescale separation may provide justification for the above statement.
\end{thebibliography}
\end{document}